\documentstyle[12pt]{article}

\input{tcilatex}
\begin{document}

\vspace*{-48pt}

\begin{center}
\noindent {\large {\bf RE-ANALYSIS OF SOME BUBBLE CHAMBER DATA ON $N%
\overline{N}$ ANNIHILATION \footnote{{\large {\bf Contribution to the
Baryons '98 Conference}}}}}

\vspace{6pt} \noindent {I.~LAZANU}\\[1mm]
\index{Lazanu, I.}{\em University of Bucharest, Faculty of Physics,
Bucharest-Magurele, Romania}\\[3mm]
\noindent {M.~RUJOIU}\\[1mm]
\index{Rujoiu, M.}{\em National Institute for Space Science, Bucharest,
Romania}\\[3mm]
\end{center}

\vspace{12pt} 
A re-analysis of some $\overline{p}p$ and $\overline{p}n$ data, at rest and
in flight, obtained in bubble chamber experiments, is presented. The ($\pi
^{+}\pi ^{-}$) and ($K_{S}K_{S}$) final states for the channels (2$\pi
^{-}\pi ^{+}$) and ($K_{S}K_{S}\pi $) are investigated. Evidence for a
narrow meson resonance structure, cautiously suggested as $f_{0}$(1500), is
given. In the ($\pi ^{+}\pi ^{-}$) invariant mass distribution from $%
\overline{p}n$ annihilations in flight, using the method of difference
spectra, a very clear evidence for $\rho ^{0}$, $f_{2}$(1270) and $f_{0}$%
(1500) is first time obtained from these data.

This re-analysis suggests that the old bubble chamber data can still provide
relevant information on the annihilation process in liquid hydrogen and
deuterium, and can elucidate controversial aspects of the annihilation
mechanism.

{\bf keywords}: nucleon-antinucleon annihilations, meson resonance

\section{Introduction}

In the last time, much of the interest in light quark spectroscopy is
dominated by the search for exotic mesons. The theoretical models predict
exotic states in the energy range 1.5 $-$ 2 GeV/c$^{2}$. The results
obtained up to now are rather ambiguous. This is due to the complexity of
the meson spectrum, where ground $q\overline{q}$ states overlap in this mass
region with radial excitations.

Gray et al.\cite{Gray}, studying the $\overline{p}n$ annihilation in the $%
3\pi $ final state, have observed a resonance in the ($\pi ^{+}\pi ^{-}$)
invariant mass distribution, with a mass of 1.525 GeV/c$^{2}$. This state is
now considered in Review of Particle Physics \cite{Rev.Part} as the $%
f_{0}(1500)$ resonance. The decay into $K\overline{K}$ final states has not
been observed in this experiment.

The possible existence of a resonant structure with a mass in the energy
range around 1.6 GeV/c$^{2}$, in the $\overline{p}p$ annihilation at rest,
in the final state $K_{S}K_{S}\pi ^{0}$, in the bubble chamber data, has
been first considered by Gastaldi \cite{Gastaldi}. For both states
considered here, the suggested quantum numbers are $I=0$ and $J=$ (even)$%
^{++}.$ The natural question is if the two structures exist and represent
the same meson, or a different one.

Scalar and tensorial mesons represent the most controversial sectors of
light mesons spectroscopy, where an excess of particle exists in respect to
the expectations derived from the naive quark model. These mesons can be
produced in $\overline{p}N$ annihilations, in association with a pion.

In the present paper we revised systematically some experimental old data
obtained in bubble chamber experiments for antiproton - nucleon
annihilations, at rest and in flight, for the channels ($3\pi $) and ($K%
\overline{K}\pi $), looking for ($\pi \pi $) and ($K\overline{K}$) signals.
The preliminary results obtained from this analysis are presented.

\section{Re-analysis of bubble chamber data}

\subsection{The channel $\overline{p}n\rightarrow 2\pi ^{-}\pi ^{+}(p)_{s}$
at rest and in flight}

The $\overline{p}n$ annihilations were obtained by antiproton deuterium -
nucleus interactions in a deuterium bubble chamber. The experimental method
selects only the annihilations on neutron, considered quasi-free if the
proton is spectator.

For this channel, the data are from references \cite{Gray} and \cite{Kalo}
for annihilations at rest, and from references \cite{bet2, Tovey} for the
process in flight.

We re-analysed these experimental data, following the method of the
so-called ''difference spectrum'', introduced by Bridges et al. \cite{Brid}.
This method eliminates the combinatorial background in the search of states
produced in $\overline{p}n$ annihilations, for the final states $\pi
^{-}X^{0}(X^{0}\rightarrow \pi ^{+}\pi ^{-})$ plus neutral. So, the
difference between $\pi ^{+}$ and $\pi ^{-}$ spectra is the spectrum of
recoiling $\pi ^{-}$, and no background is present in the absence of the
interference effects between $\pi ^{-}$ and other charged pions. The
interference effects in the low energy range decrease with the increase of
the $X^{0}$ mass, and with the decrease of the $X^{0}$ width \cite{Ze}.

The BNL group published the partial results in reference \cite{Gray}: a
sample of 2785 events obtained from 70000 $\overline{p}n$ annihilations at
rest, corresponding to proton momentum less than 150 MeV/c. The complete
analysis starts from about 3$\cdot 10^{6}$ events \cite{Kalo}. The
contamination by events in flight and/or $\pi ^{0}$ events has been
estimated to be less than 1\%. The number of all events with the proton
spectator is 5512, and includes the 2785 events from reference \cite{Gray}.
The authors consider a 0.7\% contamination of the data by non-spectators.

The invariant mass squared ($\pi \pi $) distributions recoiling $\pi ^{+}$
and $\pi ^{-}$, as well as the difference spectrum are shown in Figures 1 (a
- c). The $2\pi ^{-}\pi ^{+}$ data are dominated by $\rho $, $f_{2}(1270)$
and $f_{0}(1500)$ resonances. Large interference effects exist in this
channel; the negative values in the difference spectrum at low masses
reflect them. Because the presence of the $\rho $ resonance in the spectrum,
there are interferences at nearly all energies \cite{Ze}.

The difference spectrum fits reasonably well a superposition of $\pi
^{-}\rho $, $\pi ^{-}f_{2}(1270)$ and $\pi ^{-}f_{0}(1500)$. We used the
Breit - Wigner resonance function, and the results are in reasonably
agreement with the polynomial fits of Bridges et al. \cite{Brid} and Gray et
al. \cite{Gray}, as well as with the values of Review of Particle Physics 
\cite{Rev.Part}, in spite of the simplicity of the fit. The results of the
fit are presented in Table. 1.

The $f_{0}(1500)$ resonance was observed in the invariant mass distribution
of ($\pi ^{0}\pi ^{0})$ in the $\overline{p}p\rightarrow 3\pi ^{0}$
annihilation at rest \cite{Dev}.

For the investigation of annihilations in flight, Bettini and co-workers
collected events for the process in the energy range 1 - 1.6 GeV/c, but, in
the published paper, only the data at 1.2 GeV/c are given, which represent
818 events form 130000 pictures. The combined contribution of the
contamination of the data with events where the proton participates to
annihilations, and of the loss of good events, is estimated to be less than
2\%. In addition to this, there is an over-all scale error of $\pm $3.5\%,
obtained combining the systematic error with that of the data. The
experimental distribution for proton momentum lower than 150 MeV/c was
considered as annihilations with quasi - free neutrons.

The data collected by Tovey \cite{Tovey} represent annihilations for
incident antiproton momenta between 0.4 and 0.92 GeV/c. From a total of
150000 photographs, and after the application of all selection criteria, a
total of 2038 events were analysed.

Figures 1 (d - f) show the same distributions as in Figures 1 (a - c), but
for the reaction in flight. The number of events is 818$+$2038 and represent
the combined data from references \cite{bet2} and \cite{Tovey}, to obtain a
higher statistics. In the original analysis, no resonance behaviour has been
observed by the authors.

In the difference spectrum, structures similar with those corresponding to
annihilations at rest are visible, and we used the same fit procedure. The
results are also presented in Table 1. In the mass region 1.4 - 1.55 GeV/c$%
^{2}$, the existence of one more resonance structure is not excluded, but a
detailed analysis is not possible, due to the low statistics. For the
annihilations in flight, the fit value of the $\rho $ mass is about 40 MeV
lower than the current value. This is, probably, a combined result between
the interference effect, the rescattering effects, and the superposition of
samples obtained from annihilations in a large interval of proton momentum ($%
0.4-1.2$ GeV/c). The interference effects were before mentioned. The
rescattering effects \cite{Kob} are present in $\overline{p}n$ annihilations
as interactions of the produced mesons with the ''spectator'' proton. Due to
this rescattering, the apparent mass of the state is situated at lower
masses and larger widths. The effect increases with the increase of the
spectator proton momentum.

In the analysed sample, the annihilation events produced by antiprotons with
different momenta are considered. A comparison between the differences in $%
M^{2}(\pi \pi )$ mass distribution, for extreme cases, at rest and in
flight, is shown in Figures 2a and 2b respectively. All spectra are
normalised to unity. The difference between the ($\pi ^{+}\pi ^{-}$)
spectra, at rest and in flight, from Figure 2a, suggests that these spectra
are similar, contrary to the spectra ($\pi ^{-}\pi ^{-}$), the difference of
which is presented in Figure 2b: in the low mass region, an enhancement is
present for the annihilation at rest, that disappears for the process in
flight, where a large maximum is present at higher masses. This behaviour
dominantly affects the mass region of the $\rho $ meson.

\subsection{Neutral ($K_{S}K_{S}$) final states in the channel $\overline{p}%
N\rightarrow K\overline{K}\pi $}

Conforto et al. \cite{Conf}, and Barash et al \cite{Barash} have studied the 
$\overline{p}p$ annihilations at rest in the final states $K\overline{K}\pi $
at CERN and at BNL. The final states include the three body particles $%
K_{S}K^{\pm }\pi ^{\mp }$ and $K_{S}K_{S}\pi ^{0}$. The reaction with $%
K_{L}K_{L}\pi ^{0}$ final states is not observed in these experiments,
because the long $K_{L}$ lifetime. The combined data from these two
experiments correspond to 2.1$\cdot $10$^{6}$ $\overline{p}p$ annihilations.
Less than 10\% of the events are due to the background (one or more $\pi
^{0} $), and contaminations with annihilations in flight.

Bettini et al. \cite{bet3}, have studied the similar annihilations in liquid
deuterium in bubble chambers. From a total of 2.22$\cdot $10$^{5}$
annihilations, the final states studied are: $K_{S}K^{-}\pi ^{0}$ (140
events), $K_{S}K_{S}\pi ^{-}$ (89 events), and $K_{S}K^{0}\pi ^{-}$ (242
events).

The channel $K^{+}K^{-}\pi ^{-}$ (84 events) represents a subsample of
annihilations in which the proton spectator, with a momentum less than 250
MeV/c, is visible. A similar experiment is cited in reference \cite{Gray}.
The $K^{+}K^{-}\pi ^{-}$ sample contains 585 events, and the sample $%
K_{S}K_{S}\pi ^{-}$ 517 events respectively. Details about the experiment
are not available.

The Dalitz plots and the invariant squared projections ($K_{S}K_{S}$) and ($%
K_{S}\pi $), of the combined data from $\overline{p}p\rightarrow
K_{S}K_{S}\pi ^{0}$ and $\overline{p}n\rightarrow K_{S}K_{S}\pi ^{-}$
respectively, are shown in Figures 3 and 4. The sample $K_{S}K_{S}\pi ^{0}$
contains (182+364) events, while (84+518) events correspond to the final
state $K_{S}K_{S}\pi ^{-}$. Due to the presence of two identical particles
in the final state, the Dalitz plot are mirror symmetrised along the $%
K_{S}K_{S}$ diagonal, and contains two entries per event. In both figures,
the $K^{0*}$ and $K^{-*}$ bands are recognisable. The non-uniformity of the
distribution of events in the $K^{*}$ bands is visible. This fact could be
due to the interference effects, destructive in $K_{S}K_{S}\pi ^{0}$, and
constructive in $K_{S}K_{S}\pi ^{-}$, with diagonal band produced by $%
K_{S}K_{S}$ resonances in the energy range 1.2 - 1.6 GeV/c$^{2},$ and a
dominant $\cos ^{2}(\theta )$ distribution of the $K^{*}$ decay angle. The $%
K^{*}$ resonance is a $K\pi $ effect in the P-wave.

In the ($K_{S}\pi $) projection, the prominent structure is the $K^{*}$
signal.

We have used the fit for the $K^{*}$ resonance as a test of consistency of
adding two different samples of data.

The ($K_{S}K_{S}$) mass squared projections present complex structures. A
very weak signal is present at low masses, and corresponds to $K_{S}K_{S}$
produced at threshold. The next structures are centred in the 1250 - 1280
MeV/c$^{2}$ mass region; a very high enhancement is observed at about 1400
MeV/c$^{2},$ and the last is at 1600 MeV/c$^{2}$.

For the structure at around 1250 MeV, the assignments suggested were $%
a_{2}(1300)$, and $f_{2}(1270)$ \cite{Gastaldi, Conf}.

Between 1400 and 1500 MeV/c$^{2}$ a dip is present. If it is not due to a
negative interference effect, than the 1420 MeV band is too narrow to be
associated to the signals produced by $f_{0}(1370)$ ($\Gamma _{KK}=118-250$
MeV/c$^{2}$) or $a_{0}(1450)$ ($\Gamma =270\pm 40$ MeV/c$^{2}$), and, most
probably, it is due to the interference between different amplitudes \cite
{Amsler}.

The last structure present in this distribution, centred at about 1600 MeV,
was omitted in all initial analysis \cite{Gray, Conf, Barash, bet3}, and
first considered in reference \cite{Gastaldi}. This signal is not due to
reflections in the $K_{S}K_{S}$ projection of a $K_{S}\pi $ low energy
S-wave effect. The accumulation of events in the two bands with very low $%
K_{S}\pi $ mass at the borders of the Dalitz plot features a sudden
variation of density when crossing the orthogonal $K^{*}$ band. The density
of these low mass $K_{S}\pi $ bands is higher in the region of the Dalitz
plot common with the 1.6 GeV/c$^{2}$ band, and lower in the high $K_{S}K_{S}$
mass corner of the plot: so, the effect of high $K_{S}K_{S}$ density
superposes and interferes with the $K_{S}\pi $ S-wave effect.

This enhancement is not a kinematical reflection of the $K^{*}$ resonance.
After removing the events present in the $K^{*}$ band, comprised between 0.7
and 0.91 GeV/c$^{2}$, in the $\overline{p}p$ annihilations, about 143 events
remain in the energy window corresponding to $f_{2}(1270)$, 58 events in the
region 1400 MeV/c$^{2}$, and about 76 events in the $X(1540)$ band
respectively. For the $\overline{p}n$ annihilations, the number of events
that remain in the energetic region of interest are: 75, 89 and 129
respectively.

In the $\overline{p}p$ channel, the approximate production rate of $X(1540)$
events in the $K_{S}K_{S}\pi ^{0}$ channels is 0.32$\cdot 10^{-3}$ of all
annihilations.

The distributions of ($K_{S}K_{S}$) invariant mass squared, after the
removal of the $K^{*}$ events, are illustrated in Figures 5, a and b. For
these spectra, we performed a fit with three Breit-Wigner functions,
excluding the interference effects (the fit being also shown in the figures)
. The resonance parameters are listed in Table 2.

Another test is to verify if the 1540 MeV mass structures do not belong to
the low energy tail of a higher resonance, which decays in $K\overline{K}.$
The annihilation $\overline{p}p\rightarrow K_{S}K_{S}\pi ^{0}$ in flight has
been analysed in references \cite{Lor, Barlow} at 0.7 and 1.2 GeV/c
rsepectively.

The signal corresponding to the $X^{0\prime }(1540)$ state decreases with
the increase of the antiproton momentum..

This state is probably dominantly produced from S- and P-waves. With the
increase of the energy, the contribution of partial waves with higher
angular momentum increases, and the S- and P-waves become marginal, and this
explains the decrease of the probability of resonance production.

The final states containing charged particles are also revised. The
reactions $\overline{p}p\rightarrow K_{S}K^{\pm }\pi ^{\mp }$, with
(1897+799) events, \cite{Conf, Barash}, $\overline{p}n\rightarrow
K^{+}K^{-}\pi ^{-},$ (84+585) events \cite{Gray, bet3}, and $\overline{p}%
n\rightarrow K_{S}K^{0/-}\pi ^{-/0}$ (140+248) events \cite{bet3}, at rest,
were considered. In the $(K_{s}K^{\pm })$ effective mass squared
distribution, the accumulation of events at around 990 MeV and 1280 MeV is
clearly visible. At higher masses, no one structure is visible.

\section{Discussion of the results}

The bubble chamber data on the following annihilation processes:

$\overline{p}n\rightarrow 2\pi ^{-}\pi ^{+}$ at rest and in flight;

$\overline{p}p\rightarrow K_{S}K_{S}\pi ^{0},$ $K_{S}K^{\pm }\pi ^{\mp }$ at
rest, and

$\overline{p}n\rightarrow K_{S}K_{S}\pi ^{-},$ $K_{S}K^{0/-}\pi ^{-/0}$ at
rest

were revised.

In these preliminary results, the $\rho ^{0}$ and $f_{2}(1270)$ mesons were
identified. In the ($\pi \pi $) distribution for the (3$\pi $) final states,
for annihilations in flight, the existence of the three resonances: $\rho
^{0}$, $f_{2}(1270)$ and $X^{0}(1540)$ is first time put in evidence in the
difference spectra.

$X^{0\prime }(1540)$ was observed in ($K_{S}K_{S}$) final states.

Looking at the results obtained from this re-analysis (see Tables 1 and 2
respectively), the resonance parameters for $X^{0}$ and $X^{0\prime }$,
obtained in all reactions are consistent, and we think they characterise the
same resonance, with mean values: $M=(1541\pm 6)$ MeV and $\Gamma =(99\pm
46) $ MeV.

Because it decays into $(\pi ^{+}\pi ^{-}),$ $(\pi ^{0}\pi ^{0})$ and $%
(K_{S}K_{S}),$ and there is no evidence of a peak in the charged channels,
neither in $(\pi ^{-}\pi ^{-}),$ nor in $(K_{S}K^{\pm }),$ the isospin $I=1$
or 2 are excluded, and consequently $I=0$, with $J^{PC}$ quantum numbers
(even)$^{++}$. The simplest assumption is $0^{++}$ or $2^{++}.$

A detailed spin - parity analysis is necessary to clarify the quantum
numbers, but qualitative arguments could be given. In all investigated
channels, the production of this resonance is accompanied by the $%
f_{2}(1270).$ If $X^{0}(1540)$ and $f_{2}(1270)$ have the same spin, there
must exist a similarity of the production mechanism. But, looking to the
relative BR or production rates, $X^{0}(1540)$ does not follow the same
pattern as $f_{2}(1270),$ and we cautiously suggest $J^{PC}=0^{++},$ and in
agreement with reference \cite{Rev.Part}, the meson could be $f_{0}(1500).$

Due to the interference, most probably with $f_{0}(1370),$ the mass of the
resonance is shifted at higher values \cite{Rev.Part}. If this
interpretation is correct, it must appear in other channels. If this
supposition is not confirmed, and $J^{PC}=2^{++},$ the candidate is $%
f_{2}(1562).$

\section{Summary}

In the present paper, some data obtained in bubble chamber experiments have
been re-analysed. These experiments collected data about 30 years, in liquid
hydrogen and deuterium, in which the initial states are dominantly in S and
P - wave.

From the distributions of $(\pi \pi )$ and $(K_{S}K_{S})$ events from the $%
(3\pi )$ and $(K_{S}K_{S}\pi )$ channels, we found evidence of a narrow
resonance structure, cautiously suggested as $f_{0}(1500),$ and observed in
all investigated channels. The resonance parameters are: $M=(1541\pm 6)$ MeV
and $\Gamma =(99\pm 46)$ MeV.

In the $(\pi ^{+}\pi ^{-})$ invariant mass distribution obtained from $%
\overline{p}n$ annihilations in flight, a clear evidence for $\rho ^{0}$, $%
f_{2}(1270)$ and $f_{0}(1500)$ is obtained, using the method of difference
spectra.

In spite of the low statistics, the data from bubble chambers have a very
good quality: they permit to identify topologically the final channels, all
the detectable final states and/or decay channels can be measured with very
good precision, and the background removal is allowed. Their re-analysis
could still give relevant information and could suggest a new strategy for
the present experiments, characterised by a very high statistics.

\section{Acknowledgements}

One of the authors (I.L.) wishes to thank the organisers for invitation and
support at Baryons' 98 Conference.

\pagebreak

\begin{center}
{\Large Table 1}

Results of fits made on different spectra for 2$\pi ^{-}\pi ^{+}$ exclusive
channels in $\overline{p}n$ annihilations

\medskip

\begin{tabular}{|ccccccc|}
\hline
\multicolumn{1}{|c|}{Final state} & \multicolumn{3}{c}{at rest} & 
\multicolumn{3}{|c|}{in flight} \\ \cline{2-7}\cline{2-7}
$\pi ^{-}X$ & \multicolumn{1}{|c}{M$_{X}$} & \multicolumn{1}{|c}{$\Gamma
_{X} $} & \multicolumn{1}{|c|}{Relative} & \multicolumn{1}{|c}{M$_{X}$} & 
\multicolumn{1}{|c}{$\Gamma _{X}$} & \multicolumn{1}{|c|}{Relative} \\ 
\multicolumn{1}{|c|}{$(\rightarrow \pi ^{+}\pi ^{-})$} & (MeV/c$^{2}$) & 
\multicolumn{1}{|c}{(MeV/c$^{2}$)} & \multicolumn{1}{|c}{BR(\%)} & 
\multicolumn{1}{|c}{(MeV/c$^{2}$)} & \multicolumn{1}{|c}{(MeV/c$^{2}$)} & 
\multicolumn{1}{|c|}{BR(\%)} \\ \hline
\multicolumn{1}{|c|}{$\pi ^{-}\rho ^{0}$} & \multicolumn{1}{c|}{$772\pm 5$}
& \multicolumn{1}{c|}{$149\pm 19$} & \multicolumn{1}{c|}{$23.7$} & 
\multicolumn{1}{c|}{$727\pm 7$} & \multicolumn{1}{c|}{$104\pm 3$} & 31.1 \\ 
\multicolumn{1}{|c|}{$\pi ^{-}f_{2}$} & \multicolumn{1}{c|}{$1244\pm 4$} & 
\multicolumn{1}{c|}{$281\pm 16$} & \multicolumn{1}{c|}{71.4} & 
\multicolumn{1}{c|}{$1257\pm 1$} & \multicolumn{1}{c|}{$245\pm 26$} & 57.2
\\ 
\multicolumn{1}{|c|}{$\pi ^{-}X^{0}$} & \multicolumn{1}{c|}{$1528\pm 3$} & 
\multicolumn{1}{c|}{$69\pm 9$} & \multicolumn{1}{c|}{4.9} & 
\multicolumn{1}{c|}{$1548\pm 19$} & \multicolumn{1}{c|}{$111\pm 6$} & 11.7
\\ \hline
\end{tabular}
\end{center}

\bigskip

\bigskip

\begin{center}
{\Large Table 2}

Results of fits made for the annihilation $\overline{p}N\rightarrow
K_{S}K_{S}\pi $ at rest\medskip

\begin{tabular}{|ccccc|}
\hline
\multicolumn{1}{|c|}{Final state} & \multicolumn{2}{c}{$\overline{p}%
p\rightarrow K_{S}K_{S}\pi ^{0}$} & \multicolumn{2}{|c|}{$\overline{p}%
n\rightarrow K_{S}K_{S}\pi ^{-}$} \\ \cline{2-5}\cline{4-5}
$\pi X^{\prime }$ & \multicolumn{1}{|c}{M$_{X^{\prime }}$} & 
\multicolumn{1}{|c}{$\Gamma _{X^{\prime }}$} & \multicolumn{1}{|c}{M$_{X}$}
& \multicolumn{1}{|c|}{$\Gamma _{X}$} \\ 
\multicolumn{1}{|c|}{$(\rightarrow K_{S}K_{S})$} & (MeV/c$^{2}$) & 
\multicolumn{1}{|c}{(MeV/c$^{2}$)} & \multicolumn{1}{|c}{(MeV/c$^{2}$)} & 
\multicolumn{1}{|c|}{(MeV/c$^{2}$)} \\ \hline
\multicolumn{1}{|c|}{$\pi (f_{2}/a_{2})$} & \multicolumn{1}{c|}{$1247\pm 30$}
& \multicolumn{1}{c|}{$241\pm 66$} & \multicolumn{1}{c|}{$1265\pm 15$} & $%
155\pm 22$ \\ 
\multicolumn{1}{|c|}{$\pi X(1400)$} & \multicolumn{1}{c|}{$1394\pm 1$} & 
\multicolumn{1}{c|}{$86\pm 6$} & \multicolumn{1}{c|}{$1401\pm 4$} & $43\pm
20 $ \\ 
\multicolumn{1}{|c|}{$\pi X^{\prime 0}$} & \multicolumn{1}{c|}{$1543\pm 5$}
& \multicolumn{1}{c|}{$87\pm 23$} & \multicolumn{1}{c|}{$1544\pm 16$} & $%
130\pm 87$ \\ \hline
\end{tabular}
\end{center}

\pagebreak

\begin{center}
{\Large \ Figure Captions}
\end{center}

\smallskip Figure 1

Invariant mass squared spectra $(\pi \pi )$ and invariant squared difference
spectra for the annihilation $pn\rightarrow \pi ^{+}\pi ^{-}\pi ^{-}$ at
rest (Figure 1: a, b and c) and in flight (Figure 1 d, e and f).

(a) and (d) represents the distributions of ($\pi ^{+}\pi ^{-}$) invariant
mass squared and (b) and (e) the corresponding distributions for ($\pi
^{-}\pi ^{-}$).

(c) and (f) represent the difference spectra, as explained in the text. The
errors in the data are also shown.The fit with three Breit - Wigner
functions is superposed to the data.

\bigskip

Figure 2

(a) difference of normalised distributions between the ($\pi ^{+}\pi ^{-}$)
invariant squared masses at rest and in flight in the annihilation: $%
\overline{p}n\rightarrow \pi ^{+}\pi ^{-}\pi ^{-}.$

(b) the same for ($\pi ^{-}\pi ^{-}$).

\bigskip

Figure 3

Dalitz plot for the annihilation $\overline{p}p\rightarrow K_{S}K_{S}\pi
^{0} $ at rest, ($K_{S}K_{S}$) and ($K_{S}\pi ^{0}$) invariant mass squared
projections. The combined data from \cite{Conf}\cite{Barash} are included
(535 events). The plot is symmetrised along the $K_{S}K_{S}$ diagonal.

\bigskip

Figure 4

Dalitz plot for the annihilation $\overline{p}n\rightarrow K_{S}K_{S}\pi
^{-} $ at rest, ($K_{S}K_{S}$) and ($K_{S}\pi ^{-}$) invariant mass squared
projections. The combined data from \cite{Gray, bet3} are included.(605
events). The plot is symmetrised along the $K_{S}K_{S}$ diagonal.

\bigskip

Figure 5

$K_{S}K_{S}$ invariant mass squared for the annihilations $\overline{p}%
p\rightarrow K_{S}K_{S}\pi ^{0}$ (a), and $\overline{p}n\rightarrow
K_{S}K_{S}\pi ^{-}$ (b) at rest after the removal of the $K^{*0}$ events.
The fit with three Breit - Wigner functions is also shown.

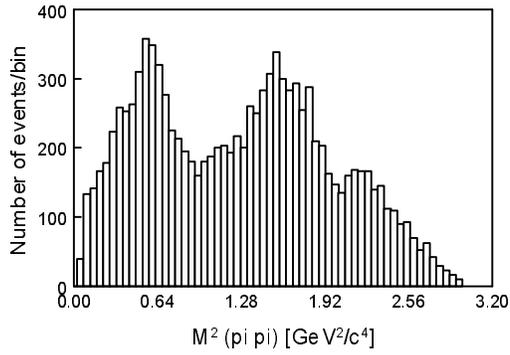
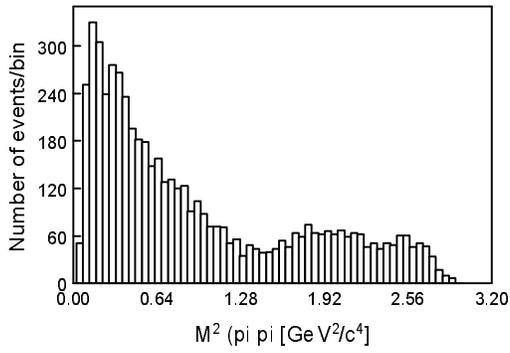
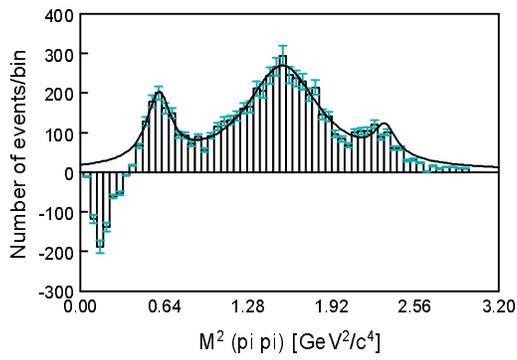
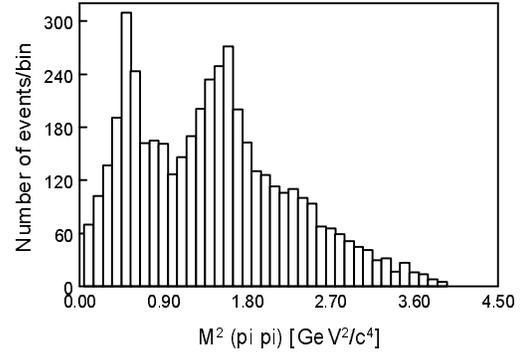
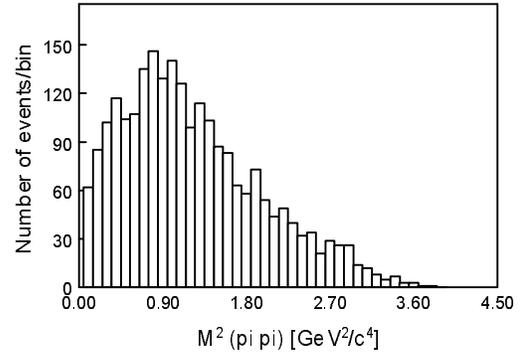
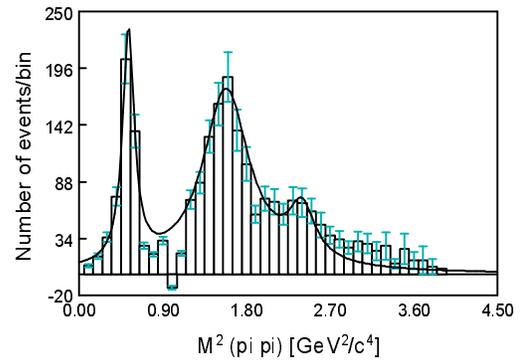
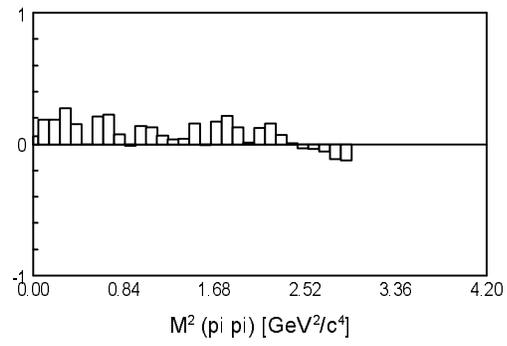
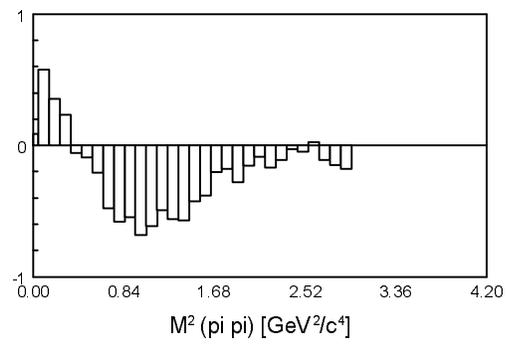
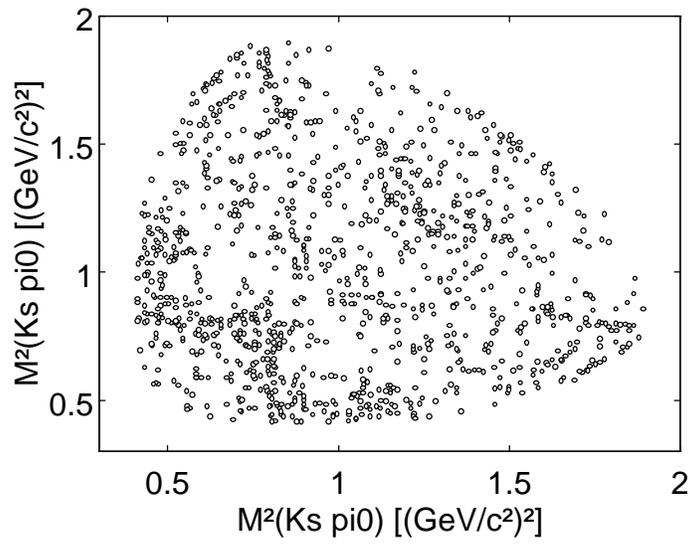
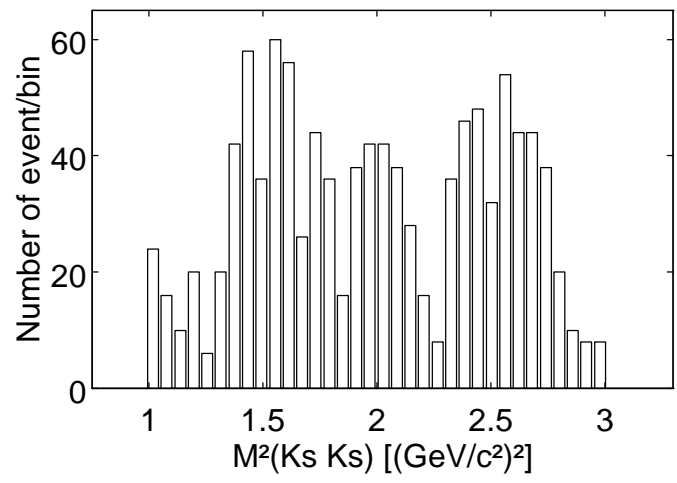
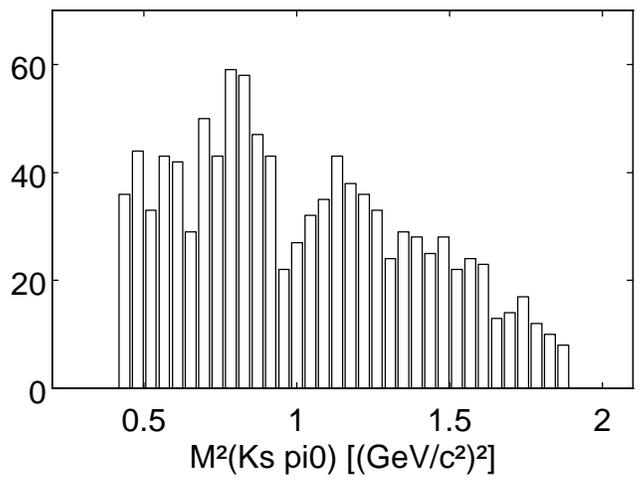
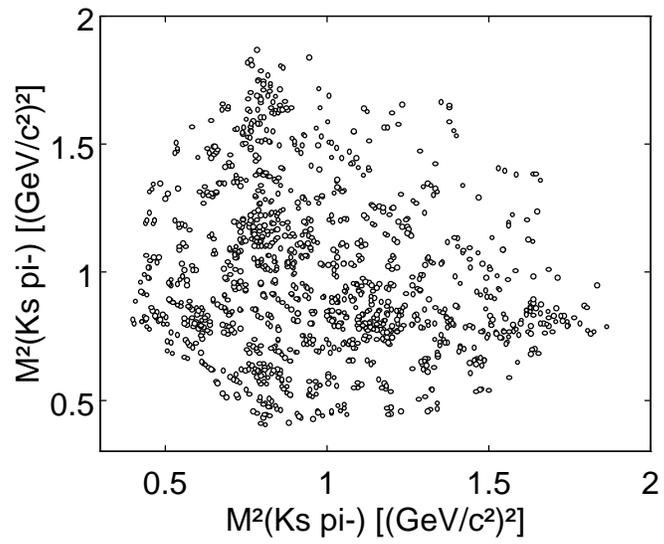
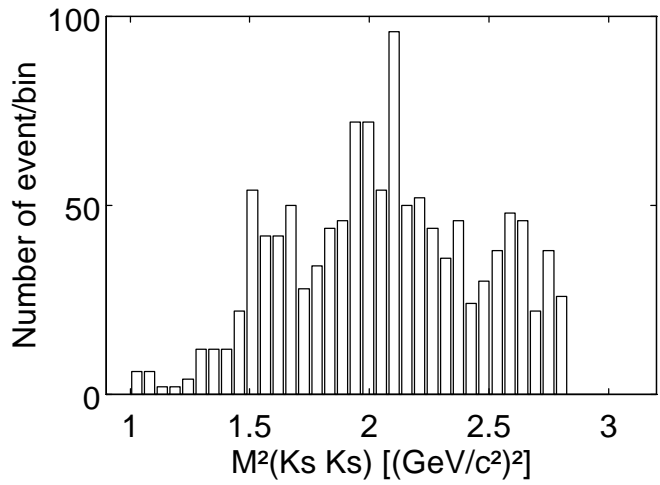
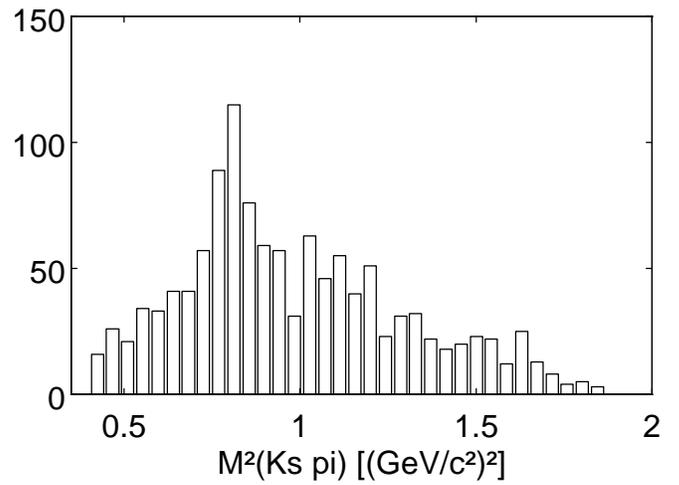
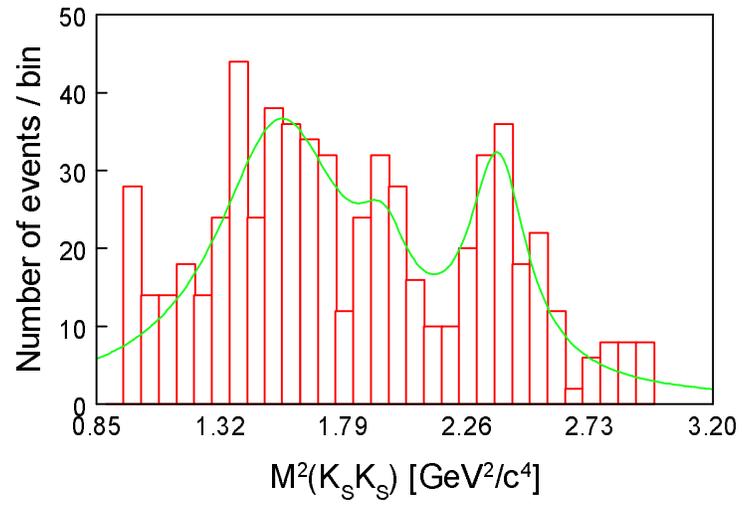
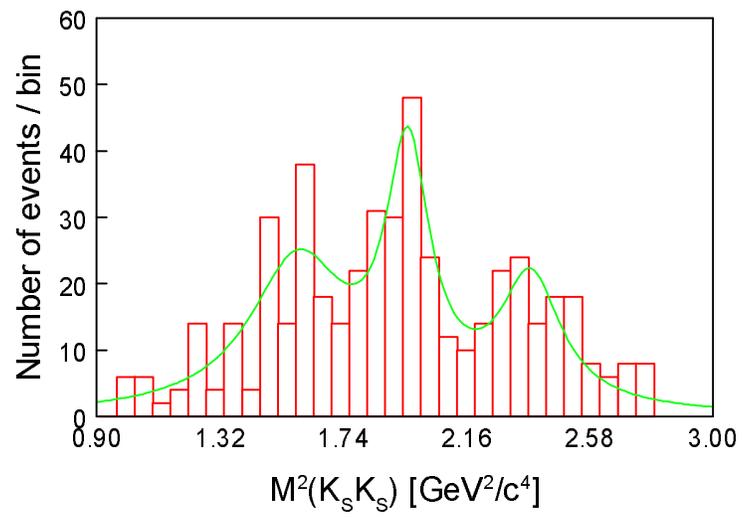
\end{document}